# Extending Polaris to Support Transactions

Josep Aguilar-Saborit, Raghu Ramakrishnan, Kevin Bocksrocker, Alan Halverson
Konstantin Kosinsky, Ryan O'Connor, Nadejda Poliakova, Moe Shafiei
Taewoo Kim, Phil Kon-Kim, Haris Mahmud-Ansari, Blazej Matuszyk, Matt Miles, Sumin Mohanan, Cristian Petculescu, Ishan Rahesh-Madan,
Emma Rose-Wirshing, Elias Yousefi

Microsoft Corp

## ABSTRACT
In Polaris [1], we introduced a cloud-native distributed query processor to perform analytics at scale. In this paper, we extend the underlying Polaris distributed computation framework, which can be thought of as a read-only transaction engine, to execute general transactions (including updates, deletes, inserts and bulk loads, in addition to queries) for Tier 1 warehousing workloads in a highly performant and predictable manner. We take advantage of the immutability of data files in log-structured data stores and build on SQL Server transaction management to deliver full transactional support with Snapshot Isolation semantics, including multi-table and multi-statement transactions. With the enhancements described in this paper, Polaris supports both query processing and transactions for T-SQL in Microsoft Fabric [2].

## 1. INTRODUCTION

*Relational data warehouses* have long been the enterprise approach to data analytics, while *lakehouses* have evolved in recent years to rapidly ingest data from many sources and give users flexible analytic tools to handle the resulting data heterogeneity and scale. A common pattern is that lakehouses are used for data preparation, and the results are then moved to a traditional warehouse to operationalize interactive analysis and reporting with high usage and SLAs. We eliminate this lakehouse/warehouse dichotomy by standardizing on a single Parquet-based [3] updatable open data format for tables in the data lake and supporting the full relational SQL tool-chain, including general CRUD transactions, directly over the diverse and large datasets stored in this standard format.

In Polaris [1], we introduced a cloud-native distributed query processor to perform analytics at scale. We observe that Polaris is really a robust *distributed computation platform* (DCP) that is designed to execute read-only queries in a scalable, dynamic and fault-tolerant way. The Polaris DCP is a stateless micro-service architecture based on (1) a careful packaging of data and processing into units called "tasks" that can be readily moved across compute nodes and re-started at the task level; (2) partitioning data into (a large number of) "cells" with a flexible distribution model; (3) a task-level "workflow-DAG" that represents inter-task dependencies efficiently; and (4) a framework for fine-grained monitoring and flexible scheduling of tasks that is resilient to failures.

In this paper, we extend the Polaris DCP to build a complete transaction manager that executes general CRUD transactions (including updates, deletes and bulk loads in addition to queries) for Tier 1 warehousing workloads.

The design is guided by the following objectives:

**1. Cloud-first, to leverage elasticity**. Transactions need to be resilient to node failures on dynamically changing topologies. For this reason, the storage engine disaggregates the source of truth for *execution state* (including data, metadata and transactional state) from compute nodes. In the cloud, data is naturally decoupled from compute nodes by using remote cloud storage. Cloud-first *transactional engines* must additionally ensure disaggregation of *metadata and transactional state* from compute nodes to ensure that the life span of a transaction is resilient to changes in the back-end compute topology, which can change dynamically to take advantage of the elastic nature of the cloud or to handle node failures. Elasticity is central to Microsoft Fabric, which is a serverless SaaS model in which customers do not need to specify cluster sizes—the system automatically allocates an appropriate number of compute resources based on the job size, and the cost is proportional to total resource consumption (i.e., #resources multiplied by time held) rather than size of cluster or number of resources allocated. For users, this means jobs in general complete faster (and usually, at less overall cost).

**2. Columnar, immutable and open storage format, optimized to handle read-heavy workloads with low contention.** Column-oriented vs. row-oriented databases have been discussed extensively, e.g., [4, 5]. A native *columnar* storage format is important to leverage the full potential of vectorized query processing for SQL, and we want to support zero-copy data sharing with other services in the lake. In addition to T-SQL, Microsoft Fabric natively supports Spark [6], data science, log-analytics, real-time ingestion and messaging, alerting, data pipelines, and reporting with Power BI. Fabric also supports interoperability with third-party services from other vendors that support the same open storage format through data virtualization mechanisms (mirroring or short-cuts). These goals led us to an open Parquet-based columnar format.

We also want to support read-heavy workloads with low contention. Traditional row or page-oriented storage engines use a write-ahead log (WAL) [5] to capture the set of actions required to undo and redo changes in data as of a point in time, i.e., to separate this information from compute nodes. This is practical for operational stores that need to support in-place data modifications for efficient ad-hoc updates in high volumes, e.g., as in TPC-C [8] type workloads. However, it is not well-suited for low-contention analytic workloads with long-running and read-intensive transactions—*immutability* of data files is important to avoid lock contention on data access. Thus, modern analytical stores based on *log-structured tables (LSTs)* such as Delta Lake [9], Iceberg [10], and Hudi [11], as well as commercial offerings from Databricks [12] and Snowflake [13][1] operate on *immutable* columnar storage

---

[1] All of these formats store data in Parquet files but differ in how they handle metadata. The Polaris internal format is similar; we currently align with Delta Lake [9] but will support APIs of all major formats through metadata converters such as Delta UniForm [27] and OneTable [26].

(Section 2) and capture changes in data via log-structured "delta" files and metadata files called *manifests*. LSTs also maintain data lineage naturally with many benefits (see below). Given these benefits, we use LSTs for table storage.

**3. Support lineage-based features (versions, time-travel, table cloning).** Traditional page-oriented storage with in-place update over a single version of the data leads to heavy data processing to reconstruct a database as of a point in time. A key benefit of LSTs is that they track data lineage natively and maintain all versions as data is updated. We take advantage of these capabilities of LSTs to efficiently support data versioning, cloning, and querying as of a point in time in Polaris, similar to other modern analytic SQL engines, including Databricks [12] and Snowflake [13].

**4. Snapshot Isolation (SI) semantics.** Analytical applications run very long-running transactions that need to execute concurrently without friction. Our transactional storage engine implements Snapshot Isolation semantics [14] over versioned data. SI allows reads and writes to proceed concurrently over their own snapshot of the data such that R/W never conflict, and W/W of active transactions only conflict if they modify the same data. As we will see in Section 4, the immutable data representation in LSTs allows us to deal with failures by simply discarding (data and metadata) files that represent uncommitted changes, similar to how we discard temporary tables during query processing failures—a key insight that allows us to leverage the Polaris DCP framework for transaction processing.

Our goal is to provide full SQL Snapshot Isolation transactional support so that all traditional data warehouse requirements are met, including multi-table and multi-statement transactions. While all mixes of transactions are supported, we note that the design is optimized for analytics, specifically read- and insert-intensive workloads—for update-intensive workloads such as TPC-C, row-oriented operational stores will be superior in performance.

**5. No cross-component state sharing**. One important architectural principle we have emphasized is *encapsulation of state within each component* to avoid sharing state across nodes. Our system, based on SI and the isolation of state across components, is designed to *execute transactions as if they were queries*, making read and write transactions indistinguishable by the Polaris DCP, thereby allowing us to fully leverage its optimized distributed execution framework.

## *1.1 Related Systems*
The most closely related systems are transaction management systems based on log-structured tables, including Big Query [15, 16], Redshift [17], Snowflake [13], Databricks [12], and open-source systems Iceberg [10], Hudi [11], Delta Lake [9], Trino [28] and Dremio [29]. In common with all these systems, we store data in immutable Parquet files and maintain table versions through "manifests" that record the changes of committed transactions. To our knowledge, Polaris is the only system that implements full Snapshot Isolation semantics, with support for multi-statement and multi-table transactions. It is also distinguished by its fully elastic SaaS model. With respect to traditional SQL databases, we are closer to warehouses, but the choice of log-structured tables represents a completely different approach.

## 2. LOG-STRUCTURED DATA STORAGE
## 2.1 Updatable Parquet Stores
Delta Lake, Iceberg and Hudi are popular open-source extensions of the Parquet format that support updates. All these storage extensions are optimized for read-heavy workloads, and to support data versioning naturally. They are variants of log-structured files —existing files are not modified in-place, and changes are represented incrementally, potentially leading to fragmentation, which is mitigated by periodic background compaction.

Intuitively, an updatable table is a collection of Parquet files. When the table is updated, new files are added and/or delete bitmaps are added to existing files, together with the metadata needed to track these changes. Subsequent access (to query or modify) the table must take these changes into account.

Log-structured files capture data changes via copy on write and/or merge on read. An insert operation always generates a new data file that is represented in the manifest via a new log entry. The difference between the two schemes is in how updates/deletes are handled:

- **Copy on write** [9] deletes the entire data file where rows are being updated and replaces it with a new file into which it copies all the contents of the previous file with new values for the updated rows.
- **Merge on read** [9, 10, 11] adds the new versions of updated rows into the new file and marks the updated rows plus any deleted rows as *deleted* (using delete bitmaps) in the existing data file, which is not modified.

Obviously, ad-hoc or trickle updates are not as efficient in comparison to traditional row stores. However, lineage of data can be easily reconstructed by mapping a commit sequence number of write transactions to changes in the manifest files. Our storage engine keeps data versioned to reconstruct the state of a table within a retention period defined by the user to support point-in-time capabilities for restore and query.

In the rest of this section, we describe log-structured table storage in detail. In Sections 3 and 4, we present our implementation of transaction management over data in this representation.

## 2.2 Storage Engine Physical Layout
In Figure 1, we show how the data in a Polaris table is organized across the *database system catalog* (logical metadata layer), a collection of *table manifests* (physical metadata layer), and a collection of files that hold the data itself. For comparison, the figure also illustrates the corresponding layers for Delta-Parquet and Iceberg formats.

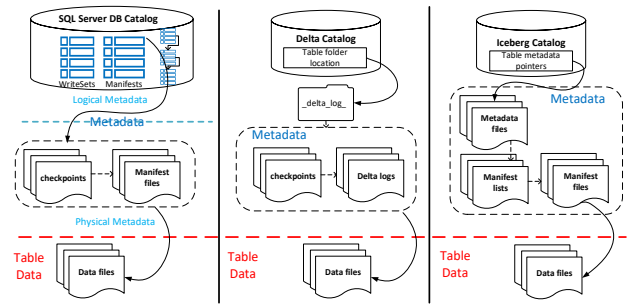

**Figure 1. Anatomy of a Log Structured Table**

This organization is best described bottom up, starting from the data files. All data files for a table in a database are grouped in a common data path in Fabric's OneLake [18]. Our internal logical and physical metadata is "published" in the form of a Delta Log table in OneLake in order to allow access by other engines such as Spark (Section 5.4). There is only one copy of the data in OneLake and publishing only involves metadata.

The lineage of the data files in a table is captured via a list of append-only log files (or *manifest files*), one per write transaction that modifies the table and commits, with sequence numbers ordered by increasing chronological order in the physical metadata layer. A sequence of manifest files defines a snapshot of the table as of a point in time and can be checkpointed for performance and memory optimization.

The sequence of manifest files and checkpoints that describe table snapshots is stored in the topmost layer, called the *logical metadata* layer. We implement this in SQL DB using a new system catalog table called the *Manifests* table. The *Manifests* table only shows files from committed transactions. A second system catalog table called *WriteSets* is used to track writes of transactions and handle conflicts, as discussed later in Section 4.

Finally, Polaris also stores the logical properties of objects within a database (tables, views, indexes, etc.) using the SQL DB catalog. Logical metadata must be sufficiently detailed for the query compiler to build high quality query plans and ensure correctness; at the same time, the access to this information needs to be very efficient since we need to tolerate high query throughput and bind many objects per query compilation. Adhering to stringent SLAs, traditional database management systems store logical metadata in system catalog databases. These databases, including SQL DB, are highly optimized for fast lookup operations and minimal contention.[2] In comparison, other table formats such as Delta and Iceberg store both logical and physical metadata in the object store layer.

## 2.3 LST: The Physical Data Layout

Data files of a table are in Parquet format that are abstracted as a collection of data *cells* that can be assigned to compute nodes to achieve parallelism. Polaris query processing operates at the cell level and is agnostic to the details of the data within a cell. Data extraction from a cell is the responsibility of the (single node) query execution engine (SQL Server) and is extensible for new data types. For scale-out processing, large datasets must be uniformly *distributed* across thousands of buckets so that they can be processed in parallel across nodes, i.e., across a large number of cells.

Parquet is a widely used open format for storing tabular data in files in a columnar form. While recent extensions consider how to handle updates (Section 3), the basic Parquet format is static. The most basic scenario is a large table stored in the lake, partitioned into a collection of Parquet files. Each Parquet data file (or a subset, if the file is large) can be mapped to a data cell.

Columnar storage by itself does not do a good job supporting indexed access to records, which is important for many workloads. We must support overlaying the columnar data with indexes if we are to be as good or better on all query workloads as traditional row-oriented, page-based architectures. We use *Z-Ordering* to support range-based retrieval over a (composite) key.

---

[2] The SQL DB catalog also stores the logical metadata associated with objects within a database. In the short term, object logical metadata is not versioned, but we plan to address this soon.

Figure 2 illustrates how a collection (e.g., table) of data objects (e.g., rows) in Polaris is logically abstracted as a collection of *cells* $C_{ij}$ containing all objects $r$ such that $p(r) = i$ and $d(r) = j$. The *distribution d(r)* is a system-defined function that returns the bucket number, or *distribution*, that $r$ belongs to. The distribution number is used to map cells to compute nodes. The *partitioning function p(r)* on the other hand is used to support range-based retrieval over a composite key. The partitioning function is defined by the order in which we sort the rows in each distribution.

Cells can be grouped physically in storage however we choose (examples of groupings are shown as dotted rectangles in Figure 1), so long as we can efficiently access $C_{ij}$. Queries can selectively reference either cell dimension or even individual cells depending on predicates and type of operations present in the query.

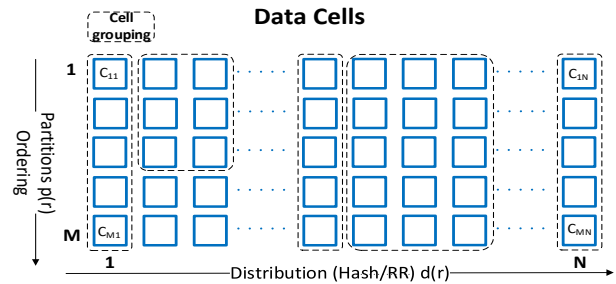

**Figure 2. Polaris Data Model**

Query processing across thousands of machines requires query resilience to node failures. For this, the data model needs to support a *flexible allocation* of cells to compute, such that upon node failure or topology change, we can re-assign cells of the lost node to the remainder of the topology. This flexible assignment of cells to compute is ensured by maintaining metadata state (specifically, the assignment of cells to compute nodes at any given time) in a durable manner outside the compute node.

## 3. POLARIS FOR TRANSACTIONS

In this section, we describe how the underlying distributed computing platform (DCP) of Polaris for query processing [1] is extended to handle full transactions with Snapshot Isolation. Figure 3 illustrates the sequence of read and write operations conducted within a user transaction. Throughout this Section, we refer to Figure 3 to reference the (steps) involved to support transactions in Polaris.

Intuitively, our design handles a distributed transaction just like a distributed query for the most part. In particular, the distributed plan comprises of (1) a SQL DB root transaction that runs at the SQL Front End (FE), and (2) sub-plans for each Back End (BE) node that leverage all optimizations for query execution, converting all inserts, updates and deletes into private changes (to both data and transaction metadata files) that are only visible to the updating transaction. The DCP aggregates changes to the transaction metadata file from the BE nodes (which use the block interface to ADLS files [19] and can proceed in parallel), and then runs the root transaction in the SQL FE, which[3] updates the logical metadata tables (see Section 4.1) using Snapshot Isolation in SQL DB. The user transaction succeeds in Polaris if the root transaction commits successfully (and is retried otherwise). In Section 4.1 we show that this ensures Snapshot Isolation over all Polaris user transactions.

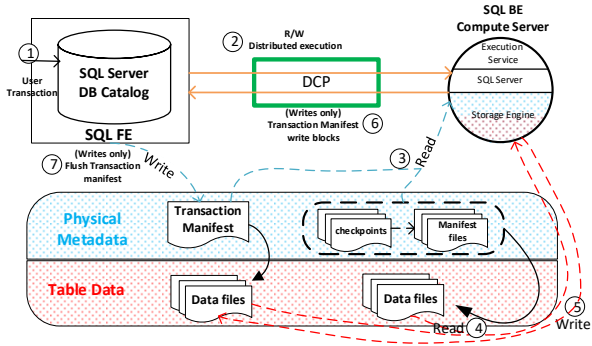

**Figure 3. Read and Write operations in Polaris.**

## 3.1 Extending the SQL DB Catalog

To manage the life cycle of a transaction in SQL FE, Polaris extends the SQL DB system catalog with two new tables per database, as we noted in Section 3. Figure 4 shows the tables (*Manifests* and *WriteSets*) and their schemas.

**Manifests**

| Table Id | ManifestFileName | Sequence Id | Transaction Id |
|---|---|---|---|
| 1001 | m0 | 0 | 100004 |
| 1202 | m1 | 1 | 100116 |

**WriteSets**

| Table Id | Updated |
|---|---|
| 1001 | 1 |
| 1202 | 0 |

**Figure 4. Manifests and WriteSets Catalog Tables**

As discussed earlier, the *Manifests* table serves as the persisted list of committed transaction manifest files for changes to any table within the database; if (transaction) *Transaction Id* modifies (table) *Table Id*, there is a corresponding row in Manifests. The *Table Id* column serves to uniquely identify objects (tables) within the database, while the S*equence ID* is the order in which (Snapshot Isolated) transactions are logically committed. The *ManifestFileName* corresponds to the name of the transaction manifest file, represented using a *guid* generated by the modifying transaction. *Transaction Id* functions as the SQL DB transaction identifier, offering durability even through service restarts. This field is required to garbage collect aborted transactions, as explained in Section 5.3.

The *WriteSets* table tracks the set of tables modified by a transaction and is used to check for WW conflicts in Polaris transactions, as we discuss later in this section.[4] The *TableId* in this table aligns with the one in the *Manifests* table, serving as a common identifier. The *Updated* field is a counter used to indicate whether (the table identified by the *TableId* column) has been updated or deleted.

Polaris leverages SQL DB transactions in the SQL FE to ensure the durability of the transaction manifests at commit. Notably, only transactions that successfully commit need to be actively tracked to ensure consistency. Any manifests and data associated with aborted transactions are systematically garbage-collected from OneLake through specialized system tasks, as elaborated in Section 5.

## 3.2 Physical Metadata and Transactions

Polaris introduces a new layer in the SQL Server storage engine, the Physical Metadata layer shown in Figure 1, to handle log-structured tables (LSTs). We now describe this layer and how it addresses read and write operations in a transaction. In Section 4, which also includes an example that illustrates read and write operations in detail, we build on this section to describe how we implement Snapshot Isolation semantics.

As described in the previous section, we capture the state of an LST table through a set of *manifest files*. Each manifest file, generated by a committed transaction, records the set of changes made by the transaction, including *data files* (Parquet) that were added or removed from the table and *deletion vectors* describing changes to individual data files. By incrementally replaying the information stored in the manifest files, the state of the table can be reconstructed.

### 3.2.1 Read Operations

The Physical Metadata layer in the SQL BE is responsible for reconstructing the table snapshot based on the set of manifest files managed in the SQL FE (3). The result is the set of Parquet data files and deletion vectors (4) that represent the snapshot of the table, and queries over these are processed by the SQL Server query execution engine.

The reconstructed state is cached in memory and organized in such a way that the table state can be efficiently reconstructed as of any point in time. This enables the cache to be used by different operations operating on different snapshots of the table. It also enables the cache to be incrementally updated as new transactions commit.

### 3.2.2 Write Operations

The Physical metadata layer in SQL BE is used to perform write operations on the log-structured table. Inserting data into a log-structured table creates a set of Parquet files (5) that are then recorded in the transaction manifest.

A transaction is represented by a single manifest file that is modified concurrently by (one or more) SQL BEs. To coordinate the concurrent writes, a SQL BE leverages the Block Blob API provided by ADLS [19]. Each SQL BE instance serializes the information about the actions it performed, either adding a Parquet file or removing it. The serialized information is then uploaded as a block to the manifest file. Uploading the block does not yet make any visible changes to the file. Each block is identified by a unique ID generated on the writing SQL BE. After completion, each SQL

---

[3] An important efficiency detail is that the SQL FE also compacts and rewrites the aggregated blocks in the transaction manifest file and delete vectors.

[4] While this paper describes conflict detection at table granularity, we discuss detection at data file granularity in Section 4.4.

BE returns the ID of the block(s) it wrote to the Polaris DCP (6). The block IDs are then aggregated by the Polaris DCP and returned to the SQL FE as the result of the query. The SQL FE further aggregates the block IDs and issues a Commit Block operation against storage with the aggregated block IDs (7). At this point, the changes to the file on storage will become effective.

Because changes to the manifest file are not visible until the Commit operation on the SQL FE, the Polaris DCP can freely restart any part of the operation in case there is a failure in the node topology. The IDs of any blocks written by previous attempts are not included in the final list of block IDs and are discarded by storage.

### 3.2.3 Explicit Transactions

Polaris supports explicit user transactions and can execute multiple statements within the same transaction in a consistent way.

The manifest file associated with the current transaction captures all the (reconciled) changes performed by the transaction. Any changes performed by prior statements in the current transaction need to be visible to any subsequent statement inside the transaction (but not outside of the transaction). For multi-statement transactions, in addition to the committed set of manifest files, the SQL BE reads the manifest file of the current transaction and then overlays these changes on the committed manifests.

For write operations, the behavior of the SQL BE depends on the type of the operation.

*Insert operations* only add new data and have no dependency on previous changes. The SQL BE can serialize the metadata blocks holding information about the newly created data files just like before. The SQL FE, instead of committing only the IDs of the blocks written by the current operation, will instead append them to the list of previously committed blocks. This effectively appends the data to the manifest file.

*Update and delete operations* are handled differently since they can potentially further modify data already modified by a prior statement in the same transaction. For example, an update operation can be followed by another update operation touching the same rows. In this case, the final transaction manifest should not contain any information about the parts from the first update that were made obsolete by the second update. The SQL BE leverages the partition assignment from the Polaris DCP to perform a distributed rewrite of the transaction manifest to reconcile the actions of the current operation with the actions recorded by the previous operations. The resulting block IDs are sent again to the SQL FE where the manifest file is committed using the (rewritten) block IDs.

## 3.3 Polaris Architecture Discussion

Earlier in this section, we described how Polaris extends SQL Server's handling of data and metadata to log-structured tables. In this section, we bring it all together to discuss the Polaris architecture for managing multi-node transactions and the key design choices.

As described in [1], Polaris is a stateless compute architecture that decouples state information from compute nodes; "state" encompasses metadata, data, transactional logs, and caches. As illustrated in Figure 5, a Polaris compute topology consists of a set of compute servers that are in essence an abstraction of a host provided by the compute fabric, each with a dedicated set of resources (disk, CPU and memory). Each compute server runs two micro-services: (a) an Execution Service (ES) that is responsible for tracking the life span of tasks assigned to a compute container by the DCP (represented as blue triangles in the Figure), and (b) a SQL Server instance that serves as the back-bone for execution of the template query for a given task and holding a cache on top of local SSDs, in addition to in-memory caching of hot data.

Over the course of extending Polaris to handle transactions in Fabric DW, we realized that separating state is necessary but not sufficient for creating a fully elastic service at a large scale. *Equally important is ensuring that state information does not cross component boundaries*. This helps reduce communication overhead, potential failure points, and the engineering complexities associated with synchronizing state information across components.

In the previous subsection, we described how the SQL Server storage engine has been extended to handle LST tables by including a physical metadata layer manager to handle manifests and data. The physical metadata state encompasses table manifests, transaction manifests, and checkpoints, all of which are externalized in OneLake just like table data. Figure 5 illustrates the extended Polaris architecture.

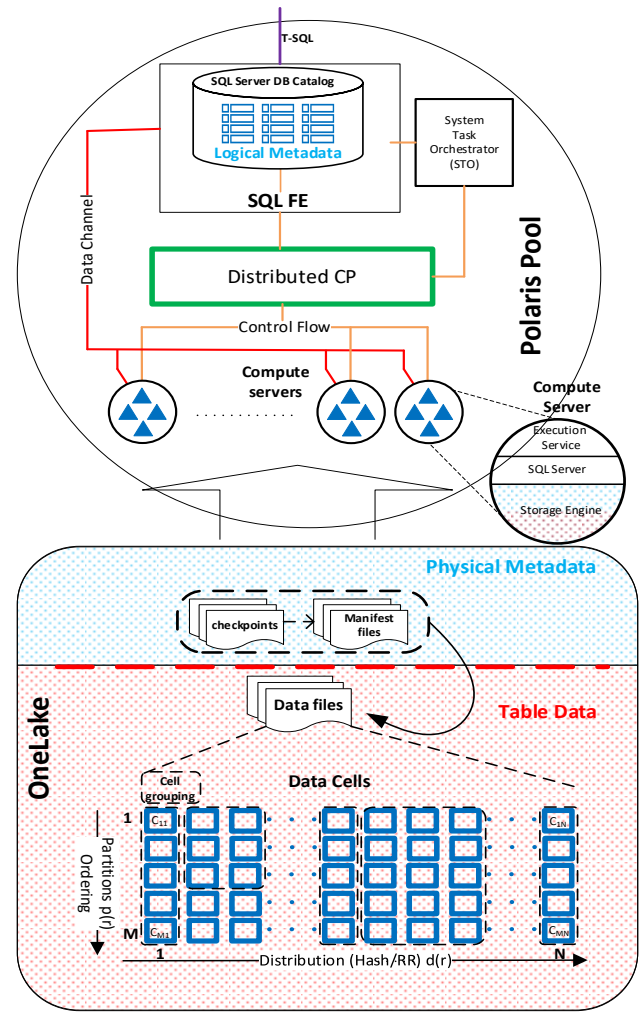

**Figure 5. Polaris Architecture**

Both physical metadata and data are cached within the SQL Server instances on the Backend (BE) nodes to enhance performance. It is important to note that the loss of these caches has no impact on the overall consistency of the system. Compute nodes can leave the topology without affecting any in-flight transactions or join the topology and replenish their caches from OneLake as queries are processed. Data can be transferred from one compute server to another via dedicated data channels. The data channel is also used by the compute servers to send results to the SQL FE that returns the results to the user. The life cycle of a query is tracked via control flow channels from the SQL FE to the DCP, and the DCP to the ES.

We use an instance of the SQL DB database service to store Polaris logical metadata as part of the SQL DB database catalog. The SQL Server Front End (SQL FE) component is responsible for compilation, authorization, authentication and transaction management. Distributed transactions in Polaris are run as user transactions within SQL FE. Utilizing SQL DB transactions provides us with important benefits, such as Snapshot Isolation semantics for logical metadata, complete compatibility with compute engine security, and the preservation of full T-SQL compatibility for both DDL and DML statements. During the life cycle of a transaction, both read and write operations are compiled in the SQL FE, leading to the generation of a distributed query plan.

Read operations follow the principles explained in [1], but with a significant architectural enhancement. In Fabric DW, we eliminate the 2-phase query optimization approach inherited from SQL DW [20]. Instead, we consolidate the query compilation stage in SQL FE. Consequently, SQL FE generates an optimized distributed query plan, serialized as a DAG to the DCP. Each node within the DAG is a subset of query operators that are instantiated as Tasks for distributed execution by the DCP. Each Task has as input a disjoint set of data cells. The unification of the compilation step in the SQL FE brings several advantages:

- It reduces the memory footprint with serialization and de-serialization of the search space (MEMO).
- It eliminates the need for a local compilation stage within BE compute nodes, enforcing a unified strategy to all BE nodes. This is crucial for eliminating the need for sharing logical metadata state in the BE nodes.

Polaris takes the distributed query plan and transforms it into a DAG with data dependency constraints, subsequently orchestrating its execution across the compute node topology. In this process, only the SQL FE is aware of transactions, while reads and writes are seamlessly handled by the DCP, which executes them as if they were regular queries. Additionally, the DCP is responsible for managing compute topology and workload management (WLM). In addition, the System Task Orchestrator (STO) plays a crucial role in managing various Storage Optimizations (Section 5), including tasks such as Data Compaction, manifest checkpointing, asynchronous manifest snapshots, and Garbage collection.

## 4. Snapshot Isolation in Polaris

In this section, we build on the earlier presentation of the Polaris architecture and data/metadata representation to describe how we do transaction management. Specifically, we ensure that all user transactions run with *Snapshot Isolation* semantics, thereby avoiding all the following anomalies: *dirty reads*, *non-repeatable reads*, and *phantoms* [14]. Read operations are never blocked and run with minimal overhead, since they look at the most recent committed (immutable) snapshot. Inserts are similarly optimized by creating new data files, not conflicting with other concurrent operations. Concurrent updates/deletes from different transactions do, however, conflict and can lead to transaction retries.

### 4.1 Optimistic MVCC Over LSTs

We now describe how we achieve Snapshot Isolation over all user transactions in Polaris using an optimistic concurrency control [21] approach, with two phases—a *Read* phase in which a transaction optimistically executes (carrying out reads and writes—in a private copy—of the database), and a *Validation* phase, in which we check for conflicts, using a variation of multi-version concurrency control [22] and either commit or retry/abort the transaction. We describe table-granularity conflict resolution through Section 4.3 and relax this to cover file-granularity conflict resolution in Section 4.4, where we also discuss how our approach can be extended gracefully to Read-Committed Snapshot Isolation and Serializability, with different tradeoffs.

Users' transactions execute in parallel across multiple BE nodes, but are managed through SQL DB transactions in the SQL FE with Snapshot Isolation semantics [23] over both the *Manifests* table and the *WriteSets* table. In effect, the *Validation* phase is centralized and builds on SQL Server's transaction management capabilities.

#### 4.1.1 Read Phase

The *read phase* of a transaction begins by capturing a snapshot of the underlying database tables. As described in Section 3.2, the snapshot is created in the SQL FE using SI over the *Manifests* table: it essentially comprises retrieving the list of visible rows within the *Manifests* table for the accessed tables, i.e., the transaction manifests for those tables written by transactions that committed before the new transaction begins.

*Read operations:* During this phase, the transaction reads from the snapshot captured above, adjusted for its own writes.

*Write operations:* The execution of an update/insert/delete DML statement in Polaris was discussed in Section 3.2 and has the following steps:

1. *Create the transaction manifest file*.
2. *Generate Parquet files or delete vector files*. Depending on the operation, either a new Parquet file for insertions or a delete vector file for deleted rows is generated. Updates are handled as a combination of deletion followed by insertion.
3. *Log pending updates*. Updates, including additions and removals, are logged into the transaction manifest file by each BE node. Each section of the manifest is compacted and flushed by the SQL FE upon success of the write operation (so subsequent statements in this transaction can use the transaction manifest file to see these changes).

#### 4.1.2 Validation Phase

The validation and commit processes are integrated into the user transaction's commit managed within the SQL FE as follows:

1. For each table that had updates or deletes executed by a write transaction, an upsert operation is issued against the *WriteSets* table.

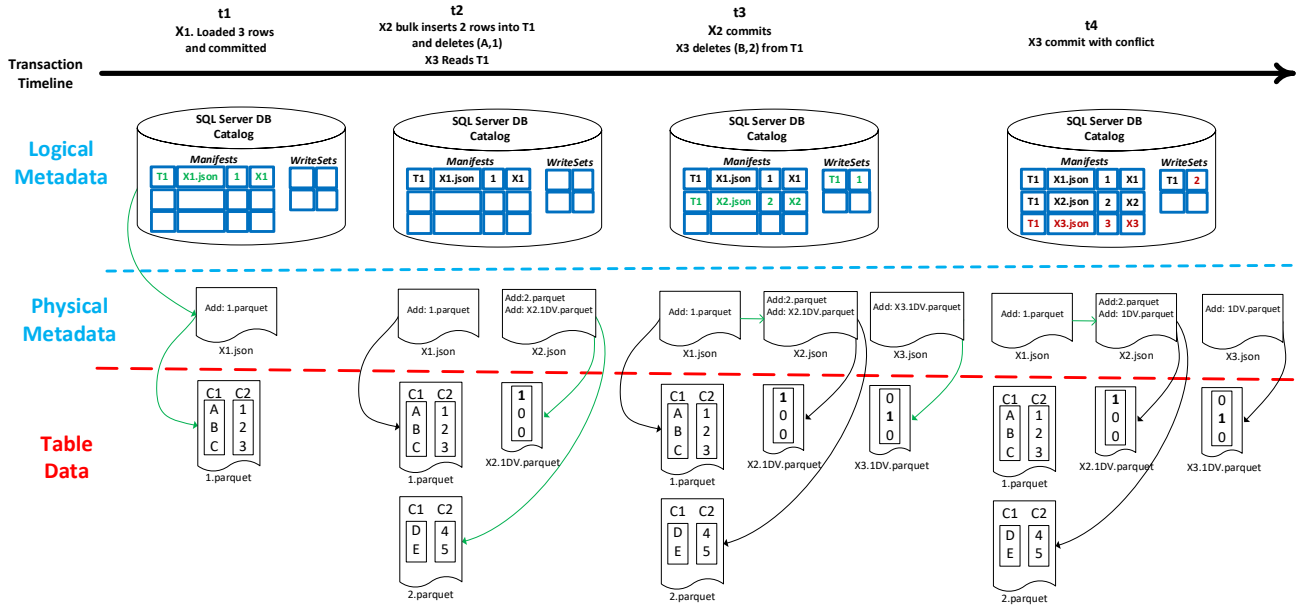

**Figure 6. Illustration of Snapshot Isolation support in Polaris**

2. A commit lock is acquired to ensure a serializable order for the transaction to be committed.
3. The transaction manifest, one for each modified table, is inserted into the *Manifests* table.
4. The user transaction is then committed within SQL FE, and the lock obtained in Step 2 is releasedThe Snapshot Isolation (SI) semantics applied to the WriteSets table plays a pivotal role in this process. It guarantees that Step 4 will fail if two concurrent transactions attempt to update the same table object, effectively resolving write-write conflicts. (We discuss conflict-resolution at data-file granularity in Section 4.4.)

In the event of a failed commit, the user transaction is rolled back, reverting changes in both the *WriteSets* and *Manifests* tables. Conversely, a successful commit makes the newly inserted manifest rows immediately visible to subsequent read and write operations. Observe that the described algorithm covers multi-table write transactions as well.

### 4.2 Example

Figure 6 illustrates the evolution of a log structured table in Polaris, T1, with columns C1 and C2, as transactions modify its contents.

***Transaction X1 (t1).*** At time t1, X1 loads and commits three rows: (A,1), (B,2), and (C,3). This operation involves writing data file1.parquet and a transaction manifest file, X1.json, to track the changes. Note that inserts are always append-only, avoiding conflicts with other transactions.

***Transactions X2 and X3 start (t2).*** At time t2, X2 inserts two rows, (D,4) and (E,5), and deletes one row (A,1); X3 reads T1. Throughout their lifespan, X2 and X3 only see the content of the Manifests table as of t2, thanks to the SI semantics in the SQL FE. For instance, a SUM(C2) query from X3 would result in 6. As for the writes, X2 generates data file 2.parquet and a new Add entry in X2.json for the inserts. Additionally, the update generates a delete vector file, 1DV.parquet, and a new entry Add into the transaction manifest. If a delete vector file was already associated with the file, two new entries would be added to the transaction manifest—one Delete for the removal of the existing delete vector and a new one Add for the merged version. In general, a single DML statement could result in the creation of many data files across several compute nodes. The transaction manifest file is atomically flushed by the SQL FE to reflect changes across all nodes when the DML statement completes, so subsequent statements in the transaction can access it. Of course, until the transaction commits and its manifest file row is entered in the Manifests table, it is not visible to other transactions.

***Transaction X2 Commit (t3).*** At time t3, X2 commits, updating the WriteSets table and inserting a row in the Manifests table to reflect the changes on T1. The SQL commit succeeds because there are no conflicting transactions.

***Transaction X3 Update (t3).*** X3 deletes row (B,2) from T1, generating a delete vector file, X3.1DV.parquet, and its corresponding Add entry in the transaction manifest. The changes made by X2 are not visible to X3 at this point, per SI semantics (because the *Manifests* table does not show the transaction manifest file for X2 yet). Thus, queries run without taking locks or being blocked by active transactions. Updates also proceed without taking locks and writing data files and transaction manifest files, optimized for bulk-writes, in particular.

***Transaction X3 Commit Attempt (t4).*** At time t4, X3 attempts to commit, upserting the WriteSets table and inserting its corresponding row in the Manifests table. However, the SQL commit detects an SI conflict in the WriteSets table, resulting in a rollback. This reverts the changes in both the Manifests and WriteSets tables.

***Potential Transaction X4 (t4).*** If a transaction X4 started at t4 and queried the SUM(C2), it would get the result 14, reflecting all actions of X1 and X2.

This detailed example illustrates Polaris's ability to handle transactions. Operating within an optimistic concurrency control framework, it strategically leverages SI semantics managed by the SQL FE, and isolates the state of active transactions from the rest of the system via their dedicated transaction manifests, enabling reads, inserts and updates to progress seamlessly without the necessity of locks. This architecture, designed for bulk-writes, also incorporates data versioning as discussed in Section 6.

## 4.3 Distributed Execution of Writes

As for read operations, the SQL FE generates the distributed plan for write operations. The primary distinction is that the root DML operation does not return data, but instead provides a list of (ADLS) block blobs in the transaction manifest file updated by the Tasks executed in the BE compute nodes. Just as with read operations, the DCP orchestrates task placement to the BE compute nodes, targeting disjoint sets of Cell IDs for write isolation.

This design ensures that the manifest entries across BE nodes do not require merging: the BE compute nodes transmit the list of blocks written for the manifest back to the DCP, which aggregates them upon completion of all Tasks before returning them to the SQL FE. The SQL FE compacts and flushes the list of block blobs upon successful completion of the write DML statement, so that the transaction manifest file is available for consumption in the BE compute nodes for subsequent statements in this transaction.

*Uniform handling of reads and writes.* Both reads and write operations in a transaction are modeled as a DAG of Tasks enabling the system to fully leverage the capabilities of the DCP, which has already been optimized for bulk writes and reads in the context of Polaris query processing to update transactions.

*Resilience to Compute Failures.* Polaris demonstrates resilience to compute failures, particularly when executing write transactions. In the event of a Task failure during a write operation, the system is capable of re-scheduling the task without causing the entire transaction to fail. The data files generated by the failed Task will be garbage collected (see Section 5.3), and the list of block blobs of the portion of the manifest reclaimed by the underlying object store, ADLS.

*Workload Separation.* Polaris employs a Workload Management (WLM) system within the DCP to effectively separate write and read workloads. Writes are isolated from reads by allocating separate sets of compute nodes for each type of operation. This separation not only improves concurrency but also prevents interference between data loading (ETL) processes and reporting workloads.

*Efficient Management of Concurrent Operations.* The Polaris system is designed to handle high concurrency effectively. Through its integrated approach, reads and writes within a transaction are seamlessly managed via Tasks, allowing for the extension of optimizations in scheduling, capacity, and resource management to be uniformly applicable to both types of operations.

## 4.4 Generalizations: Finer-Granularity Conflicts, Unique Constraints, RCSI, Serializability

We have described how Polaris supports SI semantics, but with several limitations. In this section, we address several key limitations.

### 4.4.1 Finer-Grained Conflict Resolution

For simplicity, we have presented conflict checks at the granularity of an entire table; this is reflected in the schema of the WriteSets table (Figure 4). In practice, our design allows us to check for conflicts at the granularity of data files, i.e., two concurrent updates/deletes conflict if and only if they modify the same data file (more precisely, the delete vector associated with the file, since data files are immutable). A straightforward way to check such conflicts is to extend the schema of the WriteSets table to include another column, *datafilename*. Since every user transaction's changes to WriteSets and Manifests run in a SQL DB transaction with Snapshot Isolation, if two active transactions conflict on a data file, the first to commit will succeed and the second will fail because of the corresponding row in WriteSets. (In practice, we optimize this considerably in the physical metadata layer extensions to SQL Server; we omit the details.)

### 4.4.2 RCSI and Serializable Transactions

By ensuring that all user transactions run with Snapshot Isolation semantics, we avoid all of the following anomalies, as we noted earlier: *dirty reads*, *non-repeatable reads*, and *phantoms* [14]. Our experience with SQL Server shows that users find this to be a good compromise between the overheads of serializable transactions (which SQL Server also supports) and the complexity of understanding the effects of such anomalies. Especially for analytic workloads, we believe that SI provides an excellent transaction model.

However, it must be noted that SI allows for some non-serializable interleavings of transactions [14, 24], for example:

| Transaction 1 | R(A) U(A) R(B) | C |
| Transaction 2 | | R(A) U(B) | C |

If users want serializable transactions, we can support that by running the corresponding SQL DB transaction in serializable mode [23]. The changes of the transactions (through the exposure of their manifests in the Manifests table) would then follow the serialization order.

SQL Server also supports a variant of Snapshot Isolation called Read-Committed Snapshot Isolation (RCSI) [23], wherein a transaction can read the changes of any concurrent transaction that commits. In other words, the transaction is not restricted to a snapshot as of the time it started and its own writes; it can see more recent changes affected by other transactions. We can support this mode as well by running the SQL DB transaction corresponding to the user's transaction in RCSI mode.

### 4.4.3 Unique Constraints

We do not currently enforce Unique and Primary Key constraints. To do so requires checking for duplicates, and this will have a severe impact on all changes, including inserts. In analytic scenarios, read-heavy and insert-heavy workloads are common, and such a performance hit is likely to be unacceptable.

## 5. Autonomous Storage Optimizations

The performance of Log Structured Tables is bound to fine-tuned optimizations of the underlying physical metadata. By nature, a log is additive with each transaction appending its modifications to the ever-growing log.

Performance degradation can be caused by a variety of patterns:

*Data fragmentation.* Data files can quickly become fragmented through a series of updates or deletes. The read-on-merge approach forces the SQL BE to filter out an increasing amount of data on every read.

*Small data files*. A large quantity of small (trickle-) inserts will create many Parquet files containing only a handful of rows. This amplifies the per-file IO overhead and prevents efficient compression of the data scattered across the Parquet files.

*Long list of manifests*. The list of manifests can grow on high write throughput systems, requiring the SQL BE to process an increasing amount manifest files to reconstruct the state of the table.

*Dangling files*. Aborted transactions can leave files behind on remote storage that were never committed. Likewise, files logically removed by a transaction can be deleted from storage after the retention period has expired.

The Polaris DCP relies on periodic background optimizations to identify and remedy these situations. A dedicated micro-service called the System Task Orchestrator (STO) is used to monitor the various aspects of the system. It gathers input from multiple sources and executes actions based on specific triggers. Without requiring manual user intervention, the STO can schedule optimization operations and restore read performance.

The main actions performed by the STO are explained in the following section.

## 5.1 Compaction

As part of SELECT queries, the SQL BE gathers coarse-grained statistics about the number of files and rows in the table, as well as information about files affected by deletes. These statistics are aggregated on the SQL FE and pushed to the STO.

Once a table crosses a set threshold, the STO issues a compaction operation through the SQL FE. The operation identifies the set of low-quality Parquet files, filters out any rows marked as deleted from the deletion vector and creates a new set of (compacted) Parquet files. We delve into the experimental results in Figure 10 in Section 7.3 for detailed insights.

Data compaction runs in its own transaction, executing through the SQL FE. This allows the operation to leverage the same Snapshot Isolation semantics as user transactions. Instead of deleting rewritten data files from storage, they are marked as logically removed in the manifest (for subsequent garbage collection). Likewise, the newly created data files are not visible to other transactions until the entire compaction operation is committed. The downside is that the compaction transaction can lead to unexpected conflicts with user transactions. In the future, we plan to reduce the potential for conflicts using more fine-grained conflict resolution.

## 5.2 Checkpoint

In log-structured tables each transaction adds a new manifest file that records the modifications relative to the previous set of manifests. To reconstruct the state of a table, the SQL BE needs to read every manifest and incrementally apply its changes to derive the final state. This will degrade performance as new manifests are committed over time, requiring the SQL BE to process an ever-growing list of manifest files.

The SQL FE will notify STO every time a transaction is committed. Once a table has accumulated more than a set number of manifests, STO executes a checkpoint operation over the manifests. The checkpoint operation compacts the manifest files, reconciles them, and writes out a single compacted file. The resulting checkpoint represents the full state of the table at the point in time as represented by the manifest files. We delve into the experimental results in Figure 11 in Section 7.3 for detailed insights.

Instead of reading the entire set of manifest files, the SQL BE can read the most recent checkpoint visible to the user transaction and then apply any manifest files visible to the transaction but not yet part of any checkpoint.

Checkpoints are tracked in the SQL FE through a separate table alongside the manifests. While the checkpoint operation runs in its own transaction (for Snapshot Isolation), unlike data compaction, it does not modify any data files and will not conflict with any concurrent user transactions.

## 5.3 Garbage Collection

Polaris stores both Parquet data and manifest files in a OneLake storage account. Over time, files can accumulate that are no longer referenced by the table. Aborted transactions or operations that failed (and were subsequently retried by the Polaris DCP) can leave files behind that were never committed. Data compaction or update and delete operations can mark Parquet files as logically removed. These can then be physically deleted from storage after the retention period has expired. To reclaim the space from storage, STO periodically runs a garbage collection operation.

Garbage collection reconstructs the table state based on the set of manifest files. It then sorts the data files into two sets comprising the active and inactive data files. If a data file is marked as Add it belongs to the active set. If a data file is marked as Remove and the retention period has passed, the data file belongs to the inactive set. If the file is marked as Remove but still within retention, it belongs to the active set.

In the presence of zero-copy table clone (see Section 6.2), two (or more) tables can share parts of their lineage and a single data file can be referenced by multiple tables. This requires garbage collection to process all tables with a shared lineage in one operation. Consequently, a single file can land both in the active and inactive set for different tables. In this case, the file is always considered active and removed from the inactive set if it is part of any active set.

After building the two sets, garbage collection must reconcile the list with the files on storage. Files on storage that are in the active set are reachable and are always retained. Files in the inactive set are guaranteed to be no longer reachable and can be deleted. Files on storage that are not contained in either set can belong to a currently running transaction or are leftovers from failed or aborted transactions. To distinguish them, each file is stamped on creation with the timestamp of the transaction that created it. If the timestamp is below the minimum timestamp of every currently executing transaction, then the file is guaranteed to belong to an aborted transaction and can safely be deleted. Otherwise, it needs to be retained.

## 5.4 Async Read Snapshots

Polaris provides asynchronous 'lake' snapshots for other endpoints to access DW tables. Data is already in the open and widely supported Parquet format. Without data modification, a slight transformation of the state in the physical metadata layer allows Polaris to publish table snapshots in the common open standards for Log Structured Tables: Delta, Iceberg and Hudi.

Publishing the metadata in these formats is done asynchronously through STO. As described in the previous section, the SQL FE notifies STO after each transaction commit. STO reads the committed manifest file, transforms it to the desired format and stores the transformed metadata in a user accessible location in OneLake.

By default, Polaris writes both data and manifest files into an internal location that is not accessible by the user. Data files for each table are stored in a dedicated folder in OneLake. As part of publishing a table, the underlying data folder is mapped into the user-accessible location using OneLake shortcuts, avoiding the need to copy the data. In the presence of table clones, a table can reference data files from one or more source tables, resulting in multiple shortcuts.

Currently, Polaris only supports publishing in the Delta format and the internal format of the manifest and checkpoint files closely aligns with the Delta format. Publishing a manifest file through STO as a Delta table does not require any transformation and only involves a copy of the file. Though this can be considered an implementation detail and is transparent to the user, allowing us to evolve the internal manifest format separately and add different formats in the future.

## 6. Data Lineage Specific Features

The inherent immutability of data files within Log Structured Tables provides a foundation for modern transactional engines to efficiently manage table versions across different temporal snapshots, all without the need for maintaining multiple redundant copies of the same data. Daunting operations, including Backup and Restore, traditionally requiring multiple hours, can now be executed within seconds, eliminating the need for extensive data copying. Additionally, challenging features, like querying historical versions of an object, have become straightforward through simplified metadata management of the lineage of data files. This section delves into the mechanism employed by Polaris to effectively handle data versions, unlocking capabilities that leverage this potential.

### 6.1 Query As Of

Polaris introduces a powerful feature known as *Query As Of*, allowing users to seamlessly query data as it existed at a specific point in time. Unlike traditional approaches that necessitate the creation of additional copies of the data for historical analysis, Polaris leverages the logical metadata layer in the SQL FE to time travel over the same copy of the data. This capability is achieved through the Manifests table, where the historical changes to the data are meticulously recorded. When a *Query As Of* operation is initiated, the system references the Manifests table to retrieve the relevant logical metadata for the desired time point, reconstructing a virtual snapshot of the data.

### 6.2 Clones As Of

The Manifests table enables Polaris to efficiently clone a table by only duplicating the logical metadata in the SQL FE. The clone creates a new table with the schema of the source table. The SQL FE then scans the Manifests table for every manifest recorded for the source table and re-insert the row using the Table ID of the cloned table. This effectively associates every data modification performed on the source table with the cloned table. Symmetric to Query As Of, clones can be generated as of a specific point in time by copying only the logical metadata relevant to that timeframe.

The same Snapshot Isolation semantics used for reading a table guarantee that the clone is performed in a consistent state and does not interfere with any active read or write transactions on the source table. By only inserting the manifest files created before a specific point-in-time, a clone can be created at any prior state of the table.

Since only the logical metadata is duplicated, no data (or metadata) needs to be duplicated on storage. After the clone was created, both the source and the cloned table can continue to evolve independently. New manifests can be associated with the source table without impacting the clone. Likewise, any modifications to the clone will only be associated with the clone.

### 6.3 Zero-Data Copy Backup and Restore

Building upon the idea of cloning, Backup and Restore functionalities in Polaris are optimized to work with a single copy of the data, transforming them into logical metadata only operations. Polaris secures a snapshot of all SQL Databases in the SQL FE by performing periodic Backup operations, facilitating subsequent user Restore operations of any point in time. During the restoration process to a specific point in time, the system automatically performs garbage collection on physical metadata and data files that are no longer referenced, a process explained in Section 5.3. This approach not only streamlines the Backup and Restore operations but also ensures efficient utilization of storage resources by eliminating unnecessary or outdated data.

## 7. Performance

In this section, we discuss the performance of Polaris transactions as data scales up and as the topology scales out. The first experiment shows how the system behaves as the amount of data to load into the system grows. The second shows a power run of TPC-H at the 1TB scale. We then switch our attention to LST-Bench [25] and TPC-DS queries and data for two experiments. The third demonstrates the effects of autonomous storage optimization over time as data maintenance is run in between query sessions. The final experiment covers workload isolation in the presence of concurrent read and write workloads.

All experiments were conducted using the production service compute topology in Microsoft Fabric DW. Our backend compute nodes are uniformly configured, operating within containers on physical datacenter hardware.

**REVIEWER NOTE.** *We are currently refining the ingestion and read benchmarks for TPC. Please note that the time units in the experimental results of this section are currently omitted and will be updated in the camera-ready version.*

### 7.1 Data Ingestion Scaling

In this section, we show how Polaris can provide elastic scale to support growing data scale at ingestion time.

In Figure 7 we show the load time in seconds for the TPC-H *lineitem* table at increasing scale factors of data. In our system, the maximum topology size is unbounded such that it fluctuates up and down based on compute workload demand. Note that the time increases sub-linearly with respect to the amount of data loaded. In the figure, we show the linear factor of resources used to load the *lineitem* table for that scale factor as the data label for each bar. When we load data, we estimate the cost of the load based on the amount of data, the number of source files to load from, the expected memory requirements to create high quality Parquet files, and the CPU/memory/disk requirements of the overall execution plan. In general, the CPU cost of the plan dominates, and so our parallelism is usually chosen on that basis. However, in this experiment the bottleneck is the number of files to load from. The *lineitem* table has 40 source files to read at 100GB scale and 400 source files at 1TB. We do not scale out the reading within a source file, only across source files. This results in slightly lower ingestion rates at lower data volumes and higher throughput for larger ingestion.

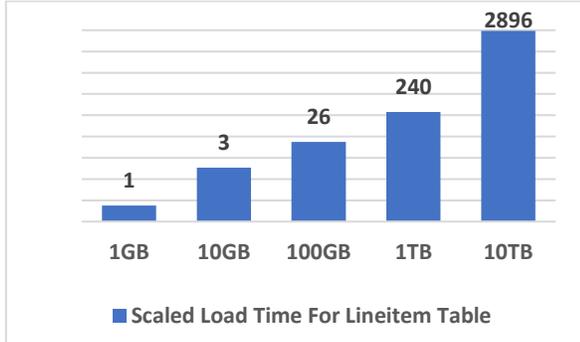

**Figure 7. Load time for the TPC-H *lineitem* table at various scale factors. Labels above the bars are the linear factor of resources used to load the data for that scale factor.**

In Figure 8, we show the total load times for 1TB and 10TB with fixed capacity as in our previous generations of the Synapse SQL DW service. As more capacity is needed to execute the load we can see the benefits of an elastic service compared to a resource-capped model. Please note that the price performance is similar since we charge for the total compute used (#resources multiplied by the total time they were used), and not the number of resources allocated. So, in Fabric DW the customer can get better performance at the same cost thanks to the serverless nature of Polaris and its cost-based resource allocation for the execution of reads and writes

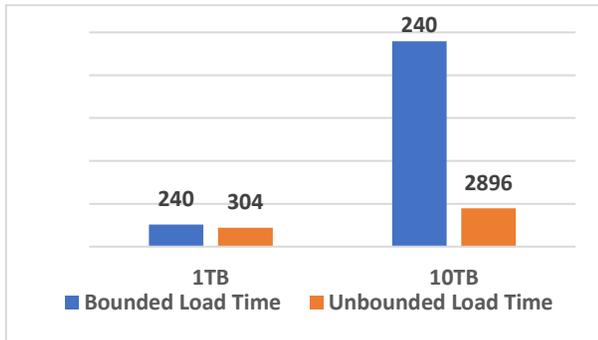

**Figure 8. *Lineitem* load times at 1TB and 10TB with fixed and elastic resources. Labels above the bars are the linear factor of resources used to load the data for that scale factor.**

## 7.2 Query Performance

We now show that Polaris continues to be an excellent distributed query engine. Figure 9 shows individual query execution times for the 22 standard TPC-H queries at the 1TB scale factor. The values presented are averages of 3 warm runs after a cold run is used to warm caches on the compute nodes. These results still hold *even when* we also run data ingestion into the same tables in parallel in a separate, uncommitted transaction. Polaris is able to isolate the load to a different set of nodes, snapshot isolation ensures that a consistent view of the data is read for all 22 queries, and caches stay warm since data files are immutable once committed.

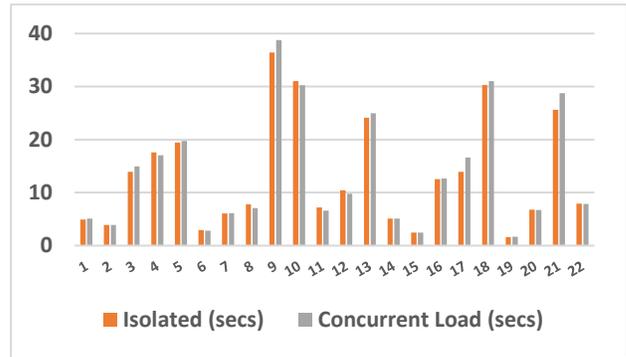

**Figure 9. TPC-H query times for 1TB scale with and without concurrent load into the same tables**

## 7.3 Autonomous Storage Optimizations

To demonstrate the effects of mixed read/write workloads, we use several LST-Bench workloads. For these experiments, we have loaded TPC-DS data at the 1TB scale and computed statistics on all columns prior to the beginning of each run.

Here we consider the effects of data maintenance on query performance, and how our autonomous storage optimizations (specifically compaction) keep query performance consistently high even when faced with large delete/update workloads.

In this experiment, we use a variant of the LST-Bench workload WP1, which alternates between a power run of the 99 TPC-DS queries (called Single User or SU in LST-Bench) and a data maintenance phase (DM). Each DM phase inserts and deletes data from the primary sales and returns tables.

Figure 10 shows the effects of autonomous discovery of data compaction during and after the DM phase, and how long the affected files stay in a suboptimal state. We show a horizontal stack of bars alternating between green and red, where green represents that all data files are within our threshold of optimality with respect to row count and deleted rows. Once DM begins, and a subsequent scan of those files occurs, we discover that the table file states have changed and report that compaction may be needed. Within a few minutes, data compaction occurs for the affected files. You can see that within a few minutes of the next SU phase starting, all tables are back to green and their storage is healthy.

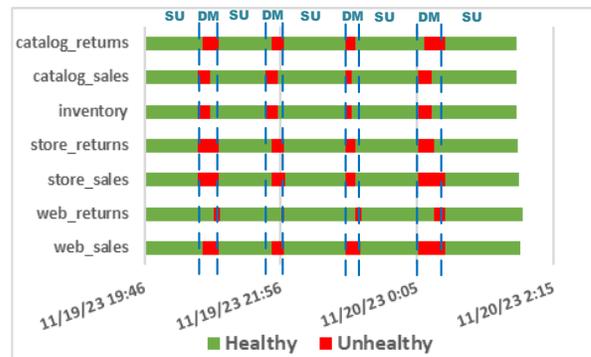

**Figure 10. Data compaction discovering and correcting storage health issues caused by WP1 data maintenance.**

In Figure 11 we show how data maintenance leads to automatic creation of manifest checkpoint files. By coincidence, each data maintenance phase creates 10 new manifest files. We run 2 INSERT statements, 6 DELETE statements, and data compaction runs twice – once between each set of 3 DELETE statements. Once we have 10 new manifest files, our checkpointing system task discovers the need for a new checkpoint and automatically creates one. The figure shows the lifetime of each checkpoint file for each table that is modified by data maintenance. For example, you can see that the catalog-related tables are modified first and the web-related tables are modified later in the DM process.

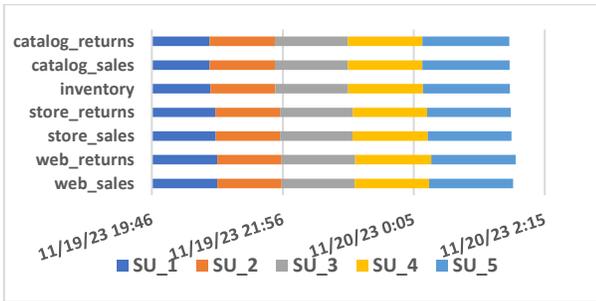

**Figure 11. Manifest checkpoint lifetimes per table within the WP1 longevity test run**

## 7.4 Read/Write Concurrency

In this section we run the LST-Bench workload called WP3, which runs concurrent Single User (SU) and Data Maintenance (DM) workloads, followed by SU and a concurrent Optimize phase. Because Polaris has autonomous storage optimization, we do not have any need to explicitly optimize storage and therefore our implementation of this benchmark runs an SU phase by itself between the phases with concurrent SU and DM.

Figure 12 shows the results of running each phase of the LST-Bench WP3 concurrency workload on the Polaris transactional engine. As expected, with concurrent data maintenance the SU phase takes significantly longer than when it runs in isolation. We expect this because the data being read is concurrently being changed by committing transactions. Snapshot isolation ensures each query in the SU stream sees a consistent view of the data, but each subsequent query gets a new snapshot based on what is currently committed. This causes delays for the same reason as in the longevity experiment – updating stats, cache misses, and committed data compaction that requires another copy of data to be read into the cache.

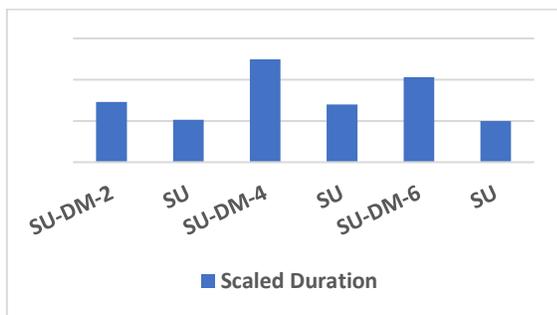

**Figure 12. LST-Bench WP3 concurrency phases**

## 8. Conclusions

This paper describes how concurrency control and recovery are handled in the SQL engine that is part of Microsoft Fabric. The approach taken is to extend the Polaris architecture, originally designed for distributed query evaluation, leveraging all its cost-based resource allocation and workload management capabilities, thereby providing a robust approach to workload isolation and resource management.

The underlying OneLake data store is based on log-structured tables with Parquet data files, with all data published continuously to open formats such as Delta Parquet—the data files remain the same, metadata files reflect differences across popular open formats.

Like other modern analytic database systems built on log-structured tables, Polaris supports time-travel, database cloning, and extremely efficient reads and inserts. Unlike other systems, a distinguishing feature is the support for full Snapshot Isolation semantics; the design can be readily extended to support Read-Committed Snapshot Isolation and serializable transactions (with corresponding performance tradeoffs).

Polaris implements automated self-healing optimizations within the system. This ensures the system's resilience and robustness by automatically identifying and rectifying potential issues or failures. The self-healing mechanisms contribute to the overall stability and reliability of the Polaris system, reducing the need for manual intervention and enhancing operational efficiency.

## 9. Acknowledgements

In this paper, we have presented transaction support in Polaris, but this represents only part of all the work that went into the SQL engine in Microsoft Fabric. Notably, it does not address the many enhancements to query processing since the original Polaris paper [1] and the extensive work that went into integrating it as a workload in the Fabric analytics suite.

We want to call out A. Vujic and N. Vujic, for their core contributions to SQL QP, VertiPaq integration, and DW platform development. Sincere thanks to S. Dash, K. Srinivasan, P. Thakkar, W. Wang, M. Kapple, and the DQP team for their pivotal roles in spearheading the integration aspects of the Polaris DCP with the transactional store. Also thanks to the rest of the Storage Engine team: S. Barlaya, X. Deng, J. Gingerich, H. Jeon, J. Ma, C. McWilliams, J. Medrano, C. Nguyen, H. Ramesh, H. Shaik, C. Watts, and H. Xu.

C. Galindo-Legaria and the Query Optimizer team did invaluable work in integrating the single-phased QO with the Polaris DQP, and C. Cunningham led our ongoing performance testing.

A. Shrinivas, P. Kumar and the entire DW infrastructure team enabled our shipping pipelines. Additionally, S. Toscano, S. Mitra, and the FE teams did outstanding work in connectivity and security, seamlessly integrating Polaris with the Microsoft Fabric platform ecosystem.

A special mention to J. Caplan, B. Crivat, R. Kaur, V. Kommineni, K. Manis, L. Mollicone, A. Netz, P. Sathy and many others who helped shape the overall product.